\documentclass[twocolumn,showpacs,preprintnumbers,prb,aps]{revtex4-1}

\usepackage{graphicx}
\usepackage{dcolumn}
\usepackage{bm}
\usepackage[section]{placeins}
\usepackage{float}
\usepackage{verbatim}
\usepackage{amsmath}
\usepackage{color}
\usepackage[colorlinks=true, letterpaper=true, pdfstartview=FitV, linkcolor=blue, citecolor=blue, urlcolor=blue]{hyperref}

\begin{document}

\title{Electric field tuning spin splitting in topological insulator quantum dots doped with a single magnetic ion}
\author{Xiaojing Li$^1$, Zhenhua Wu$^2$ and Wenkai Lou$^3$}
\email{xjli@fjnu.edu.cn}
\affiliation{$^{1}$College of Physics and Energy, Fujian Normal University, Fuzhou 350007, China}
\affiliation{$^{2}$Key Laboratory of Microelectronic Devices and Integrated Technology,
Institute of Microelectronics, Chinese Academy of Sciences, Beijing 100029,  China}
\affiliation{$^{3}$SKLSM, Institute of Semiconductors, Chinese Academy of Sciences, Beijing 100083, China}

\pacs{73.20.At;73.21.La;71.70.Ej}
\begin{abstract}
We investigate theoretically the electron spin states in disk-shaped HgTe topological insulator quantum dots (TIQDs) containing a single magnetic $Mn^{2+}$ ion. We show that the energy spectrum and the electron density distribution of the topological edge states in HgTe TIQD can be modulated significantly by the position of the magnetic $Mn^{2+}$ ion. The numerical results further demonstrate  that the electric fields not only tune the spin splittings of edge states via the $\emph{sp-d}$ exchange interaction between the electron (hole) and the magnetic $Mn^{2+}$ ion but also give rise to the bright-to-dark transitions and anti-crossing behaviors in the photoluminescence (PL) spectra. Such spin properties of HgTe TIQDs with single $Mn^{2+}$ ion as illustrated in this work could offer a new platform for topological electro-optical devices.
\end{abstract}

\maketitle

\section{Introduction }

The topological insulator (TI) is a novel class of quantum matters with insulating bulk and metallic edges or surfaces. The gapless edge or surface states are protected by time reversal symmetry (TRS) and characterized by topological invariant Z2.
The edge states or surface states of TIs
are fundamentally different from the local states caused by impurities or defects. The two-dimensional(2D) TIs were first proposed in graphene and discovered in HgTe
quantum wells (QWs) exhibiting a quantum phase
transition as the thickness of QWs ($d_{QW}$) increases.\cite{Bernevig} The
band structure of HgTe QWs is inverted when $d_{QW}>d_{c}$, i.e., the $p$-type $\Gamma_8$ band lies above the $s$-type $\Gamma_6$ band, where $d_{c}=6.3 nm$ is the critical
thickness. Gapless helical edge states in HgTe QWs which have been demonstrated experimentally via the conductance plateau when the Fermi energy is tuned inside the bulk gap.\cite{MKronig} The
formation of edge states can be understood from the gauge field arising from spin-orbit coupling.\cite{L} Besides HgTe and InAs/GaSb QWs, conventional semiconductors, e.g., InN and Ge, could also be driven into topological phases utilizing interface polarizations.\cite{M,Zhang} The band inverted QWs can produce those novel characteristics by electric field, magnetic filed and light field.\cite{Wolf}. In TIs doped with magnetic impurities, one can expect to see the interplay between the spin-orbit interactions (SOIs) and sp-d exchange interaction, which could lead to interesting phenomena, e.g., twisted RKKY interaction \cite{JiaJiZhu} and quantum anomalous Hall effect.\cite{CuiZuChang}
Comparing with bulk TIs, TI nanostructures doped with magnetic impurities have not been thoroughly studied, except for the report of HgTe well doped with $Mn^{2+}$ ions, in which magnetic moments induce an effective nonlinear Zeeman effect, causing a non-monotonic bending of the Landau levels.\cite{W}

Semiconductor quantum dots (QDs) have attracted intensive attentions in the
past decades due to their wide applications in electronic devices,
e.g., diode lasers and solar cells.\cite{Hanson,Fafard,Stanley} For various
conventional semiconductor QDs, the conduction band lies above the
valence band, i.e., normal band alignment. The electron and hole ground states are both located at the central region of the QDs,\cite{Grundmann,Hayashi,Kouwenhoven} therefore the QDs display interband optical transitions with strong oscillator strengths. By doping magnetic ions into semiconductor QDs, spin splitting and spin-relevant optical property can be changed significantly due to strong $\emph{sp-d}$ exchange interactions between electrons or holes and magnetic ions. QDs doped with a single $Mn^{2+}$ ion have been realized experimentally. \cite{Oka,Kim} The interband transition in such QDs can be observed clearly by Photoluminescence (PL) signals. \cite{Kossut,Besombes,Leger,XJLi,XJLi2} On the other hand, the disk-shaped topological insulator quantum dots have been proposed by K. Chang, et. al.\cite{KaiChang} As a consequence of quantum confinement effect, the helical edge states in a 2D TI are quantized along the circumferences of the TIQDs, leading to interesting equally-spacing edge states with a
ringlike density distribution and spin angular momentum locking. In
additional, a perpendicular magnetic field induces the persistent current
and magnetic moment of Dirac electrons which oscillate with increasing magnetic fields, ie., the Aharonov-Bohm effect.\cite{KaiChang}

In this paper we propose a disk-shaped HgTe quantum dot doped with a single $Mn^{2+}$ ion with high flexibility and controllability. We show that the energy dispersions of such TIQD doped with a single $Mn^{2+}$ ion can be effectively tuned by in-plane electric fields and the positions of the single $Mn^{2+}$ ion. The edge states of the TIQD with opposite spin orientations and angular momentum can be affected differently in a such TIQDs due to the \emph{sp-d} exchange interaction. Accordingly we illustrate the distribution of carriers in the conduction and valence bands in the presence of external electric fields and single $Mn^{2+}$ ion. Besides the energy spectra and carrier distributions, the spin-splitting of edge states can also be tuned by the electric fields and the position of doped $Mn^{2+}$ ion. Interestingly, the PL spectra of $e-h$ pairs of edge states in TIQDs exhibit spin splitting and bright-to-dark transition behaviors as the electric field and the position of $Mn^{2+}$ ion varies.

\section{Theory}

\begin{figure}
\centering
\includegraphics[width=\columnwidth]{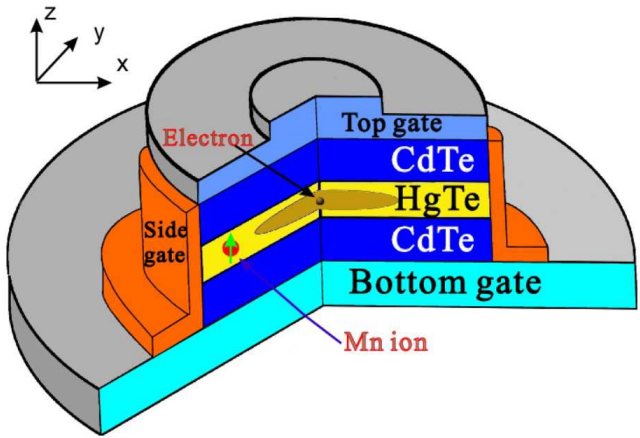}
\caption{Schematic diagram of a disk-like HgTe topological insulator quantum dot(TIQD) doped with a single $Mn^{2+}$ ion.}
\label{Fig1}
\end{figure}

We consider a HgTe QD doped with a single $Mn^{2+}$ ion as shown in Fig. \ref{Fig1}. The inverted band structure of HgTe QD in the low-energy regime can be described by so-call Bernevig-Hughes-Zhang(BHZ) model, i.e.,a four-band Hamiltonian obtained from the eight-band Kane model by neglecting the light-hole bands as shown below:
\begin{align}
H_{eff}(k_{x},k_{y})& =\left(
\begin{array}{cc}
H(k) & 0 \\
0 & H^{\ast }(-k)
\end{array}
\right)+V(\rho)+H_{sp-d} +H_{ele}\\
\end{align}
where $H(k)=\varepsilon (k)+d_{i}(k)\sigma _{i}$, $\sigma _{i}$ are the Pauli matrices,
Hermitian matrix are given by:
\begin{eqnarray}
d_{1}+id_{2} &=&A(k_{x}+ik_{y})\equiv Ak_{+} \\
d_{3} &=&M-B(k_{x}^{2}+k_{y}^{2}),   \\
\varepsilon _{k} &=&C-D(k_{x}^{2}+k_{y}^{2}).
\end{eqnarray}
where $k=(k_{x},k_{y})$ is the in-plane momentum of electrons and $A$, $B$, $C$, and $D$ are material and structure-relevant parameters. The relevant material parameters used in our calculation are $A=-0.342eV \cdot nm$, $B=-0.169eV \cdot nm^{2}$, $C=-0.00263eV$, $D=0.00514eV \cdot nm^{2}$, and $M=-0.03eV$ respectively. Note that the sign of the parameter $M$ characterizes the topological insulator phase, which is determined by the thickness of the HgTe QD. We are interested in the response of edge states to the doped $Mn^{2+}$ ion and electric field, so we take $M$ to be negative. This four-band Hamiltonian is in the band-edge Bloch basis $\left\vert
E,1/2 \right\rangle$, $\left\vert HH,3/2 \right\rangle$, $\left\vert
E,-1/2 \right\rangle$, $\left\vert HH,-3/2 \right\rangle$. $\left\vert
E\right\rangle$ ($\left\vert HH \right\rangle$)stand for electron-like(hole-like). The confining potential $V(\rho )$ of QDs can be simulated by a hard-wall potential: $V(\rho )=0,$ for $\rho <R,$ and $\infty$ for else where.$R=50nm$ is the radius of HgTe QD.

{\color{red} {In eight-band Kane model, we can consider the \emph{sp-d} exchange interaction between electron and magnetic $Mn^{2+}$ ion in the HgTe TIQD in the form:
$H_{sp-d}=-\frac{1}{2}J(r-R_{i})S_{i}\cdot \sigma$
, $J(r-R_{i})$ is the electron-ion exchange integral, $R_{i}$ is the position of $Mn^{2+}$ ion, $S_{i}$ is the total spin operator of the $Mn^{2+}$ ion at
position $R_{i}$, we can obtain the $H_{sp-d}$ of
conduction and valence band, respectively as below:
\begin{eqnarray}
(H_{s-d})_{cc} &=&\left\vert \left\langle S\right\rangle
\right\vert \alpha \overrightarrow{n}\cdot \sigma \\
(H_{p-d})_{vv} &=&\left\vert \left\langle S\right\rangle
\right\vert \beta \overrightarrow{n}\cdot \sigma
\end{eqnarray}%
where $\overrightarrow{n}=\frac{\left\langle S\right\rangle }{\left\vert
\left\langle S\right\rangle \right\vert }$, $\sigma$ is the spin operator of
the electrons at the position $r$, $\alpha$ ($\beta$) is the exchange coefficient of the electron in the conduction (valence) band with the magnetic ion. The exchange parameter of the electron in the conduction(valence) band is $\alpha=0.4eV$ $(\beta=-0.6eV )$. In our four-band model,the hole-like band $\left\vert
HH,\pm 3/2 \right\rangle$ is the heavy hole $\Gamma_{7}$, while the electron-like bands $\left\vert
E,\pm 1/2 \right\rangle$  is a linear combination of electron $\Gamma_{6}$ and light hole $\Gamma_{8}$ in eight-band Kane model. We can write the $sp-d$ interaction in the same form in the new basis.\cite{Liu} But the exchange parameters $\alpha$ should be replace by $\alpha'$ which can be given by:
\begin{eqnarray}
\alpha'=F_{1}\alpha+F_{2}\beta
\end{eqnarray}
where $F_{1}$ and $F_{2}$ is the coupling parameters depending on different well thicknesses and different Mn doping, which can be calculated from the full Kane model at small $k$.\cite{Novik,Beugeling}}} We assume that an in-plain electric field $\vec{E}$ is applied across the QD, then the electrostatic potential can be described as $H_{ele}=e\vec{E} \cdot \vec{r}$.

The eigenstates and eigenenergies can be obtained numerically by expending
the envelop wave function by using a proper set of complete basis functions as: $\Psi _{i}=\sum_{n,m}C_{n,m}^{(i)}\varphi
_{n,m}$, $C_{n,m}^{(i)}$ is the expanding coefficient. For a hard-wall circular
disk, the basic function $\varphi _{n,m}$ can be expressed as $\varphi _{n,m}=
N_{C}J_{m}(k_{n}^{m}\rho /R)e^{im\phi}$, $k_{n}^{m}$ is the $n$ the zero
point of the first kind of the cylinder Bessel function $J_{m}(x)$, $N_{C}=1/[\sqrt{\pi }RJ_{m+1}(k_{n}^{m})]$, $m=0,\pm 1,\pm 2\cdots$ is the
quantum number of the angular momentum.
After considering this exchange interaction, the new basis vector of the $H$ can be written as the direct-product of $H_{0}$ basis vector and spin state of $Mn^{2+}$ ion as $\left\vert e\right\rangle \otimes \left\vert S_{z}^{Mn}\right\rangle$ and $\left\vert hh\right\rangle \otimes \left\vert S_{z}^{Mn}\right\rangle$.

{\color{red} {The interaction Hamiltonian between the Dirac fermion and the photon within
the electrical dipole approximation is $H_{int}=H(\vec{p}+e\vec{\mathcal{A}})-H(\vec{p}%
)$~, where the vector potential $\vec{\mathcal{A}}=(\mathcal{A}_{x}\pm i\mathcal{A}_{y}%
)e^{-i\omega t}$ corresponds to the $\sigma \pm$ circularly polarized light. We suppose the fermi level locates in the middle of the bandgap.
$|i>$ denotes the initial states in the lower cones that are hole or valence
like below the fermi level, $|f>$ denotes the final states in the upper cone states that are
electron of conduction like above the fermi level. Now the electron-light interaction induces
transition from $|i>$ to $|f>$, $|i>$ and $|f>$ are written as $\Psi
_{e,h}$. The resulting optical transition rate of e-h pair
between valence and conduction band is $|<f|H_{int}|i>|$. The transition
rate is given by,
\begin{equation}
w_{if}=2\pi \delta (E_{f}-E_{i}-\hbar \omega )|<f|H_{int}|i>|^{2}.
\label{TR}
\end{equation}%
in which $<f|H_{int}|i>=\sum_{n_{1},m_{1},n_{2},m_{2}}\mathbf{C}^{+}_{f,n_{1},m_{1}}\phi_{n_{1},m_{1}}^{*}H_{int}%
\mathbf{C}_{i,n_{2},m_{2}}\phi_{i,n_{2},m_{2}}$.
}}

\section{Results and discussions}

\begin{figure}
\centering
\includegraphics[width=\columnwidth]{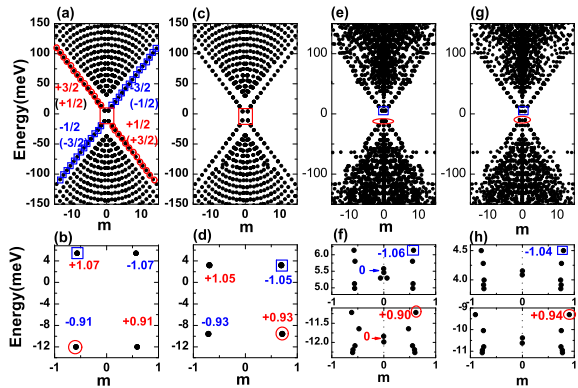}
\caption{(a)the energy spectra of HgTe TIQDs, the edge states with spin up(spin down) are denoted by the red circles(blue rectangles). The edge states and the average spin $<S_{z}>$ near the fermi surface denoted by the red square are shown in detail in (b); (c) is the same as (a), but with in-plain electric field $E=4kV/cm$ along x direction, the edge states in red square are shown in detail in (d); (e)is the same as (a), but with a single $Mn^{2+}$ magnetic ion located at the edge of the TIQD, the edge states and corresponding average spin $<S_{z}>$ in red and blue circles are shown in detail in (f); (g) is the energy spectra when the in-plain electric field and a single $Mn^{2+}$ magnetic ion located at the edge of the TIQD exist, the edge states and corresponding average spin $<S_{z}>$ denoted in red and blue circles are shown in detail in (h).}
\label{Fig2}
\end{figure}

First we illustrate the energy spectrum of TIQD changes with in-plain electric fields and the position of doped $Mn^{2+}$ ion in Fig. \ref{Fig2}. The edge states of TIQD show an approximately linear dependence on the angular momentum $m$ near the Dirac point in Fig. \ref{Fig2}(a), and the energy spectrum is gapped around $18meV$. The spin-up (spin-down) edge states are denoted by the red circles (blue rectangles). Notice that the dominant spin elements of edge states in the conduction band are $\pm 3/2 $ (i.e., the $HH$), while the dominant spin components of edge states in valence band are $\pm 1/2 $ (i.e., the $E$), which arises from the band inversion of HgTe. From the edge state near Fermi surface denoted by the red square in Fig. \ref{Fig2}(b), we can find the dominant spin component of edge state is $-3/2$($+3/2$) when angular quantum number $m$ is positive(negative). It means that electrons with opposite spin orientations propagate along opposite directions, i.e., the spin-angular momentum locking. Apparently the average spins of the edge states are mixed states due to the off-diagonal elements in the Hamiltonian coupling the $\left\vert
E,1/2 \right\rangle$ and $\left\vert HH,3/2 \right\rangle$( or $\left\vert
E,-1/2 \right\rangle$ and $\left\vert HH,-3/2 \right\rangle$). We calculate the magnitude of average spin in \ref{Fig2}(b). When we apply an in-plane electric field along the x direction, both the edge states and bulk states are affected significantly by the electric field as shown in Fig. \ref{Fig2}(c). The in-plain electric fields change the energy spectra of TIQD in two different ways. Due to the Stark effect, the electric field decreases the bandgap of the TIQD. The gap of edge states decreases to $13meV$, compared to the $18meV$ bandgap when no electric field is applied. as a result, the electric field changes the average spin of the edge states by increasing the coupling the electrons and heavy-hole states.(see Fig. \ref{Fig2}(d))

\begin{figure}
\centering
\includegraphics[width=\columnwidth]{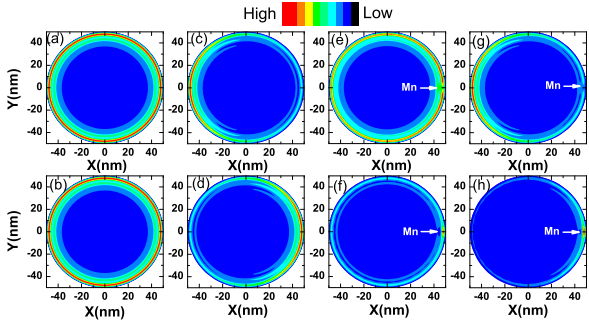}
\caption{(a) and (b) denote the density distributions of the edge states in the bottom of conduction band and on the top of the valence band marked by the blue and red circles in Fig.2(b) respectively; (c) and (d) denote the states shown blue and red circles in Fig.2(d) with an in-plain electric field; (e) and (f) denote the states shown blue and red circles in Fig.2(f) and (g) with a $Mn^{2+}$ magnetic ion located at the edge; (g) and (h) show the distributions of the edge states with both in-plane electric field and $Mn^{2+}$ magnetic ion located at the edge.}
\label{Fig3}
\end{figure}

\begin{figure}
\centering
\includegraphics[width=\columnwidth]{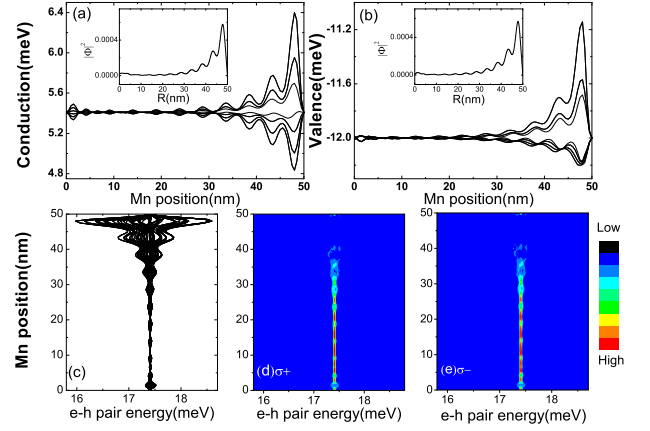}
\caption{(a) and (b): The energies of the edge states from the conduction and valence bands as a function of the position of $Mn^{2+}$ ion; the insets in (a) and (b) denote the density distributions of edge states in conduction and valence band, respectively. (c): The energies of the lowest several levels of the e-h pairs of TIQDs as a function of the position of $Mn^{2+}$ ion; (d) and (e): The photoluminescence spectra of the lowest e-h pair as a function of the position of $Mn^{2+}$ ion for the right and left polarized light $\sigma\pm$.}
\label{Fig4}
\end{figure}

Next we discuss the effects of the $Mn^{2+}$ ion located at the edge of the TIQD($R=47nm$) on the energy spectrum of TIQD in Fig. \ref{Fig2}(e). This is because that the $sp-d$ interaction between $Mn^{2+}$ ion and the edge states are maximum and the spin splitting is the largest for the edge states (see Fig. \ref{Fig4}). The existence of $Mn^{2+}$ ion breaks the time-reversal symmetry and rotation symmetry of the TIQD, and the total angular momentum is no longer a good quantum number. In this case, we still label the different states by their average angular momentum. We can see that the edge states can still exist when the angular momentum $m$ is small. It is robust against the low density of magnetic impurity. The edge states in the bottom of conduction band and the ones on top of valence band are shown in detail in Fig. \ref{Fig2}(f). The energies of edge states split vertically and horizontally. We notice the edge states in the conduction (or valence) bands are both  splitting into twelve states corresponding to the coupling states between hole-like spin and $Mn^{2+}$ spin(electron-like spin and $Mn^{2+}$ spin). The splitting of energy spectrum arises from the giant zeeman effect, i.e., the diagonal elements of the $s-d$ and $p-d$ interaction. The states which are more closer to angular momentum $0$ mix strongly with other spin states. The average spins for the edge states when angular momentums approach $0$  is almost vanish, which means the probability is almost the same for $+ 1/2$ spin and $-1/2$ spin in valence band (or $+ 3/2$ spin and $-3/2$ spin in conduction band). The horizontal shift of energy spectrum comes from the off-diagonal elements of the $s-d$ interaction, inducing the coupling between the states s $\left\vert e, Sz \right\rangle$ and $\left\vert e\pm 1, Sz\mp 1 \right\rangle$ via simultaneous spin flip of the the $Mn^{2+}$ ion and electrons. The horizontal shift is larger for electrons in valence band than that in conduction band. Because the main spin component in valence band is  $\pm 1/2$ which can couple with the $Mn^{2+}$ ion through off-diagonal elements while the electrons in
conduction band with main spin $\pm 3/2$  can not flip with the $Mn^{2+}$ spin
. Due to the $Mn^{2+}$ ion, the maximum spin splitting of edge states in conduction band or valence band is nearly $1meV$. We apply an in-plain electric field along the x direction in such TIQD doped with $Mn^{2+}$ ion in Fig. \ref{Fig2}(g). We enlarge the spin splitting in Fig. \ref{Fig2} (h). We can find the spin splitting for edge states in valence band increases to $2meV$, while the spin slitting for edge states in conduction decrease to $0.5meV$. This is because the in-plain electric field push the wavefunction of the electron of the edge states, resulting in changing the overlap the electrons and magnetic ion. We detail the discussion in Fig. \ref{Fig3}. We can tune the spin splitting of the edge states through the external electric field. Simultaneously, the angular momentum shifts of edge states can also be affected by the electric field.

The spin splitting changed by the electric field can be understood from the density distributions of the edge states in the bottom the conduction band and on top of the valence band as shown in Fig. \ref{Fig3}. In Fig. \ref{Fig3}(a) and (b), we plot the density distributions of edge states of the TIQD denoted by blue rectangle and red circle in Fig. \ref{Fig2}(b)(the edge states from conduction and valence band) which show ring-like distribution. When an in-plane electric field is applied, it pushes the distribution of edge state in the bottom of conduction band to the left in Fig. \ref{Fig3}(c), while push the distribution of edge state on the top of valence band to the opposite direction as shown in Fig. \ref{Fig3}(d). When we dope a $Mn^{2+}$ ion on the edge of TIQD, the density distribution of edge state denoted by the blue rectangle and red circle in Fig. \ref{Fig2}(f) are shown in Fig. \ref{Fig3}(e) and (f). The distribution of edge state in conduction band is still ring-like but is localized away from the position of $Mn^{2+}$ ion. The spins orientation are the same for electrons and the $Mn^{+2}$ ion. Due to the  antiferromagnetic \emph{s-d} exchange interaction, there is a repel interaction between the electron and the $Mn^{+2}$ ion with the same spin orientation. While the density distribution of edge state on top of the valence band shows opposite location which is close to the $Mn^{2+}$ ion in Figs. \ref{Fig3}(f). It arises from the opposite spin orientations between electrons in valence band and the $Mn^{+2}$ ion. Due to the ferromagnetic \emph{p-d} exchange, it results in an attractive force between the electrons in the valence band and the $Mn^{+2}$ ion. The $Mn^{2+}$ ion behaves like a scattering center for the edge states. When we consider an in-plane electric field applied across such $Mn^{2+}$ ion doped TIQD, we can see the electric field push the edge state in conduction band to the left side away from the $Mn^{2+}$ ion, while it push the edge state in valence band to the right side close to the $Mn^{2+}$ ion in Figs. \ref{Fig3}(g) and (h), respectively. Spin splittings of edge states due to the $sp-d$ interaction in the Hamiltonian between electron and $Mn^{2+}$ ion is a contact interaction depending on the overlap between the wavefunctions of electron/hole and $Mn^{2+}$ ion. So the in-plane electric field will increase the overlap of wavefunction and strength of $p-d$ interaction, resulting the larger spin splitting in valence band as shown in lower figure in Fig. \ref{Fig2} (h), while it decreases the $s-d$ interaction and the spin splittings in conduction band as shown in the upper panel of Fig. \ref{Fig2} (h).

Due to the $sp-d$ interaction, we plot the energies of the edge states in the bottom of conduction band(electron) and on the top of valence band(hole) as a function of the position of $Mn^{2+}$ ion in Figs. \ref{Fig4}(a) and (b). We find that the energies of the edge states stay almost the same when the position of $Mn^{2+}$ ion varies from $0$ to $30nm$ in both conduction and valence bands, but change abruptly when the magnetic ion locates near the boundary where the edge states locate. The \emph{sp-d} interaction between the edge states and magnetic ion depends on the overlap between the wavefunctions of electron/hole and $Mn^{2+}$ ion. Hence, there is almost no interaction between the edge states and $Mn^{2+}$ ion when the magnetic ion located at the center of TIQD. Notably, the energy spectrum of edge states near the bottom of conduction band shows opposite dependence against to that near the top of valence band. It arises from that the opposite signs of the \emph{s-d} and \emph{p-d} exchange interactions. The energy gap between the conduction and valence bands becomes smaller due to the \emph{sp-d} exchange interaction between the  electron and $Mn^{2+}$ with the same spin orientation. We also plot the higher energy levels of electron and hole, a larger energy gap appears due to the \emph{sp-d} exchange interaction between the electron and $Mn^{2+}$ with the opposite spin orientation. Notice there are several vibration of energy and spin splitting from $0$ to $30nm$. It is the result of the wavefunction vibration of electron and hole edge states (see the distribution of electron and hole in the inset of Figs. \ref{Fig4}(a) and (b)). We displays the lowest several electron-hole (e-h) pair energy of TIQDs as the function of the position of $Mn^{2+}$ ion in Fig. \ref{Fig4}(c). When the $Mn^{2+}$ ion is located away from the edge of the TIQD, the energy spectrum of e-h pair are degenerate because of weak interaction between the $Mn^{2+}$ ion and the edge stats. As the $Mn^{2+}$ ion moves towards the edge of the TIQD, the strong $sp-d$ interaction splits the energy levels and lifts the spin degeneracy. Notice the energies of the e-h pair states oscillate arising from the oscillation of the wavefunction of the ground edge state in the TIQD, which lead to the oscillating behavior of spin splitting. Finally the photoluminescence(PL) spectra for the right and left polarized light $\sigma\pm$ are plotted in Figs. \ref{Fig4} (d) and \ref{Fig4} (e). When the $Mn^{2+}$ ion locates in the center of the TIQD, the absorption spectrum with right and left polarized light are almost degenerate and bright. We notice that the bright-to-dark transitions occur when the $Mn^{2+}$ ion moves from the center($r=0$) to the edge($r=30nm$). This feature arises from the oscillation of the  wavefunction of edge states, which make the strength of $sp-d$ and spin splitting oscillates. The $s-d$ and $p-d$ interactions have opposite interaction on electrons and holes, which make the overlap of electron wavefunction from the conduction and valence bands decrease(see the distributions in Figs. \ref{Fig3}(e) and \ref{Fig3}(f)), resulting in the dark transition in spectra. When the $Mn^{2+}$ ion moves to the edge of the QD. The conduction and valence bands are strongly coupled by the $sp-d$ interaction. There is a bright-to-dark transition when the $Mn^{2+}$ ion is located around 40nm in the spectra for both $\sigma\pm$.  When the $Mn^{2+}$ ion is located at the edge of the TIQD, the wave-function overlap between conduction and valence band decreases(see Fig.3(e) and (f)), resulting in the relatively dark in the PL spectra. The e-h pair energy and PL behavior is quite different from the conventional semiconductor QD, in which the maximum spin splitting between electrons and the magnetic ion happens when the the magnetic ion locates in the center of the QD. The distinct signatures of magnetic impurity on the edge states discussed above could be possibly observed in STM experiments.

\begin{figure}
\centering
\includegraphics[width=\columnwidth]{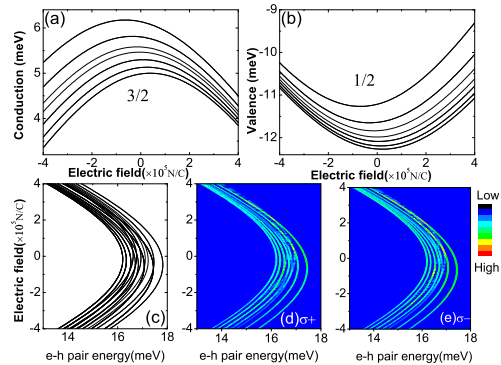}
\caption{(a) and (b): The energies of the edge states from the conduction and valence bands as the function of an in-plane electric field in the TIQD doped with a $Mn^{2+}$ magnetic ion is located at the edge of TIQD ($R=47nm$); (c) displays the lowest  e-h pair energies of TIQDs as the function of in-plane electric field; The photoluminescence spectra of the e-h pair energy as a function of in-plane electric field for right and left polarized light $\sigma\pm$ are shown in (d) and (e).}
\label{Fig5}
\end{figure}

The spin splittings of edge states become maximum when the $Mn^{2+}$ ion is located near the boundary when an in-plane electric field is applied across the TIQD. The energy spectrum of the edge states near the bottom of the conduction band and on the top of valence band changes significantly in Figs. \ref{Fig5}(a) and (b). When the electric field increases along the x direction, the energies for edge states in the conduction band decrease, while that in the valence band increase duo to the Stark effect in Fig. \ref{Fig2}. The spin splittings for the edge states in the conduction band decrease while that in the valence band increase as the electric field increases. Because the in-plane electric field pushes the wavefunctions of edge states in the conduction band with spin component $\pm3/2$ away from the $Mn^{2+}$ ion as shown in Fig. \ref{Fig3}(c) resulting in the reduced spin splittings, while it pushes the wavefunction of the edge states in the valence band with spin component $\pm1/2$ towards the $Mn^{2+}$ ion as shown in Fig. \ref{Fig3}(d) with larger spin splittings. When the electric field is applied in the opposite direction. we can find the opposite change of the energy spectrum and spin splittings. By the application of electric field, we can realize tuning spin splittings of the edge states. We also plot the energy spectrum of e-h pair in Fig. \ref{Fig5}(c). We can find the same trend of spin-splitting as the edge states in conduction band. Notice that the maximum spin splittings of lowest e-h pair states appear when the electric field is small. It is the competition between the $p-d$ splitting of the edge states near the bottom of the conduction band and the $s-d$ splitting that near the top of the valence band, because the edge states near the CBM and VBM are hole-like and electron-like due to the band inversion, respectively. Apparently the hole-like edge states in conduction band play more important role in it with electric field applied along x direction. In the PL spectrum in Fig. \ref{Fig4}(d) and (e), we can not see the spin splitting when $Mn^{2+}$ ion is located near the boundary because it is relatively weak compared with the one when the $Mn^{2+}$ ion is away from the edge states. Here we adjust the brightness to view the spin splitting in PL spectrum in \ref{Fig5}(d) and (e). We can find the spin splitting varying with the in-plain electric field. In the PL spectrum we can find the anisotropic PL spectrum for $\sigma\pm$ light. It arises from the unsymmetrical $sp-d$ elements in Four-band Hamiltonian that can only couple the electron-like states, but have no coupling between the hole-like states. The spin splittings can be detected with increasing in-plane electric fields. Here, we propose a tunable spin splitting in topological edge states using the in-plane external electric field. The TIQD doped with $Mn^{2+}$ ion offer us a promising platform for potential application in a electric control of spin-splitting topological device.

\section{Conclusions}
In summary, we demonstrate theoretically the energy spectrum and the spin splittings in the disk-like HgTe QDs containing a single $Mn^{2+}$ ion. We find that the energy spectrum and distribution of electrons in the TIQDs can be manipulated significantly by the position of $Mn^{2+}$ ion and in-plane electric fields, so as the coupling between the edge states and $Mn^{2+}$ ion and the bandgap. The edge states in the conduction band and valence band show opposite density distributions with in-plane the electric field. The spin splitting due to the $sp-d$ exchange interaction can also be tuned by the in-plane electric field. In PL spectra, the spin splittings and bright-to-dark transition can be seen by changing the position of $Mn^{2+}$ ion and electric field.  This electrical switching behavior offers us an efficient way to control the edge states which is robust against the local $Mn^{2+}$ ion doping and pave a way to construct the topological photo-electronic device. Our results provide a unique platform to understand and manipulate the electric and optical properties of edge states in TIQDs.

\begin{acknowledgments}
This work was financially supported by .....
\end{acknowledgments}


\begin{thebibliography}{99}
\bibitem{Bernevig} Bernevig, B. A. \emph{et al}. Quantum spin Hall effect and topological phase transition in HgTe quantum wells. Science, \textbf{314} 1757 (2006).

\bibitem{MKronig} Kronig,M. \emph{et al}. Quantum spin Hall insulator state in HgTe quantum wells. Science \textbf{318},766 (2007).

\bibitem{L}Shi, L. K. \emph{et al}. Anomalous Electron Trajectory in Topological Insulators. Phys. Rev. B, \textbf{87}, 161115 (2013).

\bibitem{M}Miao, M. S. \emph{et al}. Polarization-driven topological insulator transition in a GaN/InN/GaN quantum well. Phys. Rev. Lett, \textbf{109}, 186803 (2012).

\bibitem{Zhang} Zhang, D.  \emph{et al}. Interface-induced topological insulator transition in GaAs/Ge/GaAs quantum wells. Phys. Rev. Lett, \textbf{111}, 156402 (2013).

\bibitem{Wolf} Wolf, S. A. \emph{et al}. Spintronics: a spin-based electronics vision for the future. Science, \textbf{294} 1488 (2001).

\bibitem{JiaJiZhu} Zhu, J. J \emph{et al}. Electrically controllable surface magnetism on the surface of topological insulators. Phys. Rev. Lett, \textbf{106} 097201 (2011).

\bibitem{CuiZuChang} Chang, C. Z.  \emph{et al}. Experimental observation of the quantum anomalous Hall effect in a magnetic topological insulator. Science, \textbf{340} 167 (2013).

\bibitem{W}Beugeling, W. \emph{et al}. Reentrant topological phases in Mn-doped HgTe quantum wells. Phys. Rev. B \textbf{85}, 195304 (2012).

\bibitem{Hanson} Hanson, R. \emph{et al}. Spins in few-electron quantum dots. Rev. Mod. Phys, \textbf{79} 1217 (2007).

\bibitem{Fafard} Fafard, S. \emph{et al}. Red-emitting semiconductor quantum dot lasers. Science, \textbf{274} 1350 (1996).

\bibitem{Stanley} Marto, A. \emph{et al}. Production of photocurrent due to intermediate-to-conduction-band transitions: a demonstration of a key operating principle of the intermediate-band solar cell. Phys.Rev. Lett, \textbf{97} 247701 (2006).

\bibitem{Grundmann} Grundmann, M. \emph{et al}. Ultranarrow luminescence lines from single quantum dots. Phys. Rev. Lett, \textbf{74} 4043 (1995).

\bibitem{Hayashi} Hayashi, T. \emph{et al}. Coherent manipulation of electronic states in a double quantum dot. Phys. Rev. Lett, \textbf{91} 226804 (2003).

\bibitem{Kouwenhoven} Kouwenhoven, L. P. \emph{et al}. Excitation Spectra of Circular, Few-Electron Quantum Dots. Science, \textbf{278} 1788 (1997).

\bibitem{Oka} Oka, Y. \emph{et al}. Dynamics of excitonic magnetic polarons in nanostructure diluted magnetic semiconductors. J. Lumin. \textbf{83}, 83 (1999).

\bibitem{Kim} Kim, C. S. \emph{et al}. CdSe quantum dots in a $Zn_{1- x}Mn_{x}Se$ matrix: new effects due to the presence of Mn.
J. Cryst. Growth \textbf{214}, 395 (2000).

\bibitem{Kossut}Kossut, J. \emph{et al}. Cathodoluminescence study of diluted magnetic semiconductor quantum well/micromagnet hybrid structures.  Appl. Phys. Lett.
\textbf{79}, 1789 (2001).

\bibitem{Besombes}Besombes, L. \emph{et al}. Probing the spin state of a single magnetic ion in an individual quantum dot. Phys. Rev. Lett. \textbf{93}, 207403 (2004);

\bibitem{Leger} Besombes, L. \emph{et al}. Carrier-induced spin splitting of an individual magnetic atom embedded in a quantum dot. Phys. Rev. B \textbf{71}, 161307(R) (2005).

\bibitem{XJLi} Li, X. J  \& Chang, K. Electric-field tuning s-d exchange interaction in quantum dots. Appl. Phys. Lett. \textbf{92}, 071116 (2008);

\bibitem{XJLi2} Li, X. J  \& Chang, K. Electric-field switching of exciton spin splitting in coupled quantum dots. Appl. Phys. Lett. \textbf{92} 251114 (2008).

\bibitem{KaiChang} Chang, K. \& Lou, W. K. Helical quantum states in HgTe quantum dots with inverted band structures. Phys. Rev. Lett, \textbf{72} 106,206802 (2011).

\bibitem{Liu} Qin Liu and Tianxing Ma. Classical spins in topological insulators. Phys. Rev. B,\textbf{80} 115216 (2009).

\bibitem{Novik} Novik, E. G. \emph{et al}. Band structure of semimagnetic $Hg_{1-y}Mn_{y}Te$ quantum wells. Phys. Rev. B,\textbf{72} 035321 (2005).

\bibitem{Beugeling} W. Beugeling, C. X. Liu, E. G. Novik, L. W. Molenkamp, and C. Morais Smith. Reentrant topological phases in Mn-doped HgTe quantum wells. Phys. Rev. B,\textbf{85} 195304 (2012).


\end{thebibliography}
\end{document}